\documentclass[aps,preprint]{revtex4}%
\usepackage{amsfonts}
\usepackage{amsmath}
\usepackage{amssymb}
\usepackage{graphicx}%
\providecommand{\U}[1]{\protect\rule{.1in}{.1in}}

\begin{document}
\title{A comparative study of the electronic and magnetic properties of BaFe$_{2}%
$As$_{2}$ and BaMn$_{2}$As$_{2}$ using the Gutzwiller approximation}
\author{Y. X. Yao}
\affiliation{Ames Laboratory-U.S. DOE. and Department of Physics and Astronomy, Iowa State
University, Ames, Iowa 50011, USA}
\author{J. Schmalian}
\affiliation{Ames Laboratory-U.S. DOE. and Department of Physics and Astronomy, Iowa State
University, Ames, Iowa 50011, USA}
\affiliation{Karlsruhe Institute of Technology, Institute for Theory of Condensed Matter,
76131 Karlsruhe, Germany}
\author{C. Z. Wang, K. M. Ho}
\affiliation{Ames Laboratory-U.S. DOE. and Department of Physics and Astronomy, Iowa State
University, Ames, Iowa 50011, USA }
\author{G. Kotliar}
\affiliation{Department of Physics and Astronomy, Rutgers University, Piscataway, New
Jersey 08854, USA}

\begin{abstract}
To elucidate the role played by the transition metal ion in the pnictide
materials, we compare the electronic and magnetic properties of BaFe$_{2}%
$As$_{2}$ with BaMn$_{2}$As$_{2}$. To this end we employ the LDA+Gutzwiller
method to analyze the mass renormalizations and the size of the ordered
magnetic moment of the two systems. We study a model that contains all five
transition metal 3d orbitals together with the Ba-5d and As-4p states
(ddp-model) and compare these results with a downfolded model that consists of
Fe/Mn d-states only (d-model). Electronic correlations are treated using the
multiband Gutzwiller approximation. The paramagnetic phase has also been
investigated using LDA+Gutzwiller method with electron density
self-consistency. The renormalization factors for the correlated Mn 3d
orbitals in the paramagnetic phase of BaMn$_{2}$As$_{2}$ are shown to be
generally smaller than those of BaFe$_{2}$As$_{2}$, which indicates that
BaMn$_{2}$As$_{2}$ has stronger electron correlation effect than BaFe$_{2}%
$As$_{2}$. The screening effect of the main As 4p electrons to the correlated
Fe/Mn 3d electrons is evident by the systematic shift of the results to larger
Hund's rule coupling J side from the ddp-model compared with those from the
d-model. A gradual transition from paramagnetic state to the antiferromagnetic
ground state with increasing $J$\ is obtained for the models of BaFe$_{2}%
$As$_{2}$ which has a small experimental magnetic moment; while a rather sharp
jump occurs for the models of BaMn$_{2}$As$_{2}$, which has a large
experimental magnetic moment. The key difference between the two systems is
shown to be the d-level occupation. BaMn$_{2}$As$_{2}$, with approximately
five d-electrons per Mn atom, is for same values of the electron correlations
closer to the transition to a Mott insulating state than BaFe$_{2}$As$_{2}$.
Here an orbitally selective Mott transition, required for a system with close
to six electrons only occurs at significantly larger values for the Coulomb interactions.

\end{abstract}

\maketitle

\section{Introduction}

The discovery of high T$_{c}$ superconductivity in LaO$_{1-x}$F$_{x}$FeAs has
initiated a detailed investigation of these and related iron based compounds
with the ultimate aim to identify new high temperature
superconductors\cite{Kamihara08}. Many binary, ternary and quaternary
compounds have been synthesized and investigated experimentally, some of which
were quickly found to exhibit high T$_{c}$ superconductivity with electron or
hole doping or under pressure or intrinsically
\cite{Rotter08,Wu08,Tapp08,Torika08,Chen08,Ren08}. Meanwhile, several closely
related and physically interesting systems have also been studied. An
interesting example is the BaMn$_{2}$As$_{2}$ compound, which has the same
layered tetragonal ThCr$_{2}$Si$_{2}$-type crystal structure as the BaFe$_{2}%
$As$_{2}$ at room temperature\cite{Singh09,Singh091,An09,Johnston10}.
BaMn$_{2}$As$_{2}$ is an antiferromagnetic (AFM) insulator with a small gap;
while BaFe$_{2}$As$_{2}$, a parent compound of the Fe-based high T$_{c}$
superconductors, is an AFM metal at low temperatures. The more localized
behavior of the ordered magnetic state in the Mn-based system suggests that
electronic correlations in this system are stronger or more efficient in
changing the electronic properties of the material. A closer, comparative
investigation of distinct transition metal ions therefore promises to reveal
further whether iron is indeed special among its neighbors in the periodic
table. This is also expected on qualitative grounds on the basis of the Hund's
rule picture\cite{Haule09}.

Theoretically, the description of the normal states for iron pnictide
superconductors is complicated by the multi-orbital nature and the presence of
electron correlation effects. Interesting results can be obtained from low
energy theories that are primarily concerned with states in the immediate
vicinity of the Fermi
surface\cite{Kuroki08,Raghu08,Chubukov08,Cvetkovic09,Zhai09,Sknepnek09}.
However, material specific insight often requires a careful analysis of states
over a larger energy window. In the iron pinctide and chalcogenides the
strength of the correlations is controlled by the Hund's rule coupling $J$,
rather than by the Hubbard $U$ or the p-d charge transfer energy\cite{Haule09}%
. This makes these materials a new class of correlated metals, different from
doped Mott or charge transfer insulators. Calculations based on density
functional theory (DFT) usually yield too wide Fe 3d bands and too large
magnetic moment in the AFM ground state\cite{Mazin08}. Recently the AFM phase
of the iron pnictide parent compounds has been successfully described by using
a combination of DFT and dynamical mean field theory (DMFT) with full charge
self-consistency\cite{Yin11}.\ These investigations clearly demonstrated the
important role played by the Hund's rule coupling $J$ for the magnetic and
electronic properties of the iron pnictides. A systematic analysis of a large
class of Fe-based materials further demonstrated that the nature and strength
of local correlations are rather universal, while material specifics result
from changes in the kinetic energy, amplified by the frustrated nature of
electron hopping path due to Fe-As-Fe and direct Fe-Fe overlaps\cite{Yin11b}.
This universality in the value of the local Coulomb interactions $U$,
$U^{\prime}$, $J$ only exists if one includes in addition to the Fe-3d states
at least the As 4p orbitals. Downfolding to models with iron states only is
sufficient to lead to material specific variations in the local correlations.
In view of these insights is it important to compare the behavior of Fe-based
systems with systems based on other transition metals, also in view of the
ongoing debate whether these systems should be considered weakly or strongly correlated.

On a technical level, quite a few efforts have been devoted to study the
electronic and magnetic properties of the systems using the Gutzwiller
variational method\cite{Dai10} and simplified tight-binding
models\cite{zhou10,Bune11} with the particular appeal that the Gutzwiller
approach is computationally cheaper if compared to more sophisticated DMFT
approaches such as the usage of Quantum Monte Carlo algorithms. For example.
the experimental magnetic moment was obtained in a three band Hubbard model of
LaOFeAs which contains only Fe 3d $t_{2g}$ orbitals\cite{zhou10}. Calculations
based on the five-band model show that the $e_{g}$ orbitals are also important
and the system exhibits a sharp transition from paramagnetic (PM) state to AFM
state with large magnetic moment when $J$\ increases, which is different from
the results of three-band model\cite{Bune11}. Another important question is
whether, in addition to the Fe 3d $e_{g}$ orbitals, the inclusion of As 4p
elections is required for a more quantitative description of the AFM and
paramagnetic phases. Finally, recent calculations for LaOFeAs \ based on the
LDA+Gutzwiller approach that includes a Fe3d--As4p Wannier-orbital basis does
indeed reproduce the experimentally observed small ordered magnetic moment
over a large region of ($U,J$) parameter space\cite{Weber}.

In this paper, we report a comparative study of the BaFe$_{2}$As$_{2}$ and
BaMn$_{2}$As$_{2}$\ on the electronic and magnetic properties by investigating
the Fe/Mn d-band model and the ddp-model that includes Fe/Mn 3d, Ba 5d, and As
4p orbitals with Gutzwiller
approximation\cite{Gutzwiller1,Gutzwiller2,Bunemann07} which has been shown to
be equivalent to a slave-boson mean field theory\cite{Kotliar86}. The
paramagnetic phase for the two systems has also been investigated using
LDA+Gutzwiller method with electron density self-consistency\cite{Dai08,Dai09}%
. BaMn$_{2}$As$_{2}$ is shown to be a more correlated material than BaFe$_{2}%
$As$_{2}$. The screening effect of the As 4p electrons is evident by the
comparison between the results of the Fe/Mn d-model and those of the ddp-model
or LDA+Gutzwiller method. The key difference in the behavior of these systems
is shown to be the distinct electron count of the 3d states, a result that is
robust against changes in the detailed description of these systems. While the
even number of electrons for an Fe site requires that the system undergoes an
orbitally selective Mott transition, Mn with five electrons can more easily
localize and therefore has a much stronger tendency towards large moment
magnetism and Mott localization, even for the nominally same values of the
Coulomb interactions. The prospect to tune the tendency towards localization
by varying the transition metal ions may therefore play an important role in
optimizing the strength of electronic correlations for high temperature superconductivity.

\section{Formalism}

For the sake of being self-contained, we outline the main formalisms for the
multiband model with the Gutzwiller
approximation\cite{Gutzwiller1,Gutzwiller2,Bunemann07} and the LDA+Gutzwiller
approach with electron density self-consistency\cite{Dai08,Dai09}.

\subsection{Multiband model with the Gutzwiller approximation}

The multi-band Hamiltonian with local many-body interactions is written as%
\begin{equation}
\mathcal{H}=\mathcal{H}_{0}+\mathcal{H}_{1} \label{Hamil}%
\end{equation}
where the bare paramagnetic band Hamiltonian ($\mathcal{H}_{0}$)\ is%
\begin{equation}
\mathcal{H}_{0}=\sum_{\left(  i\alpha\right)  \neq\left(  j\beta\right)
,\sigma}t_{i\alpha j\beta}c_{i\alpha\sigma}^{\dagger}c_{j\beta\sigma}%
+\sum_{i,\alpha,\sigma}\varepsilon_{i\alpha}c_{i\alpha\sigma}^{\dagger
}c_{i\alpha\sigma}%
\end{equation}
$t_{i\alpha j\beta}$ is the electron hopping element between orbital
$\varphi_{\alpha}$ at site $i$ and orbital $\varphi_{\beta}$ at site $j$.
$\varepsilon_{i\alpha}$ is the orbital level.$\ $For the Fe/Mn\ d-model,
$\alpha$ and $\beta$\ run over the Fe/Mn 3d orbitals. For the ddp-model,
$\alpha$ and $\beta$\ run over the Fe/Mn 3d orbitals, Ba 5d orbitals, and As
4p orbitals. $c^{\dagger}$($c$) is the electron creation (annihilation)
operator. $\sigma$\ is the spin index.

The typical Hubbard density-density type interaction term ($\mathcal{H}_{1}$)
for Fe 3d electrons is%
\begin{align}
\mathcal{H}_{1}  &  =U\sum_{i,\gamma}n_{i\gamma\uparrow}n_{i\gamma\downarrow
}+U^{\prime}\sum_{i,\gamma<\gamma^{\prime},\sigma\sigma^{\prime}}%
n_{i\gamma\sigma}n_{i\gamma^{\prime}\sigma^{\prime}}\nonumber\\
&  -J\sum_{i,\gamma<\gamma^{\prime},\sigma}n_{i\gamma\sigma}n_{i\gamma
^{\prime}\sigma}%
\end{align}
which is usually sufficient for describing collinear magnetic
order\cite{zhou10}. Here $\gamma$\ is the correlated Fe/Mn 3d orbital index,
$U^{\prime}=U-2J$, and $n_{i\gamma\sigma}=c_{i\gamma\sigma}^{\dagger
}c_{i\gamma\sigma}$.

We take a variational wave function of the Gutzwiller form,%
\begin{equation}
\left\vert \Psi_{G}\right\rangle =\frac{\hat{G}\left\vert \Psi_{0}%
\right\rangle }{\sqrt{\left\langle \Psi_{0}\left\vert \hat{G}^{2}\right\vert
\Psi_{0}\right\rangle }}%
\end{equation}
with the Gutzwiller approximation to calculate the expectation values of Eq.
\ref{Hamil} \cite{Gutzwiller1,Gutzwiller2,Bunemann07}. $\Psi_{0}$ is the
uncorrelated wave function. $\hat{G}$ is the Gutzwiller projection operator,
which adopts the following form%
\begin{equation}
\hat{G}=e^{-\sum_{i\mathcal{F}}g_{i\mathcal{F}}\left\vert \mathcal{F}%
_{i}\right\rangle \left\langle \mathcal{F}_{i}\right\vert }%
\end{equation}
where $\left\vert \mathcal{F}_{i}\right\rangle $ is the Fock state generated
by a set of $\left\{  c_{i\gamma\sigma}^{\dagger}\right\}  $%
\begin{equation}
\left\vert \mathcal{F}_{i}\right\rangle =%
{\displaystyle\prod\limits_{\gamma\sigma}}
\left(  c_{i\gamma\sigma}^{\dagger}\right)  ^{n_{i\gamma\sigma}^{\mathcal{F}}%
}\left\vert 0\right\rangle
\end{equation}
with $n_{i\gamma\sigma}^{\mathcal{F}}=0$ or $1$, which identifies whether
there is an electron with spin $\sigma$ occupied in orbital $\gamma$ for Fock
state $\mathcal{F}$ at the $i^{th}$\ site. $g_{i\mathcal{F}}$\ is a
variational parameter which controls the weight of the local Fock state
$\left\vert \mathcal{F}_{i}\right\rangle $ in the Gutzwiller wave function.
$g_{i\mathcal{F}}=1$ for the empty and singly occupied configurations because
in these cases there are no electron-electron repulsion involved. According to
Ref. \cite{Bunemann07}, the expectation value of the electron Hamiltonian
$\mathcal{H}$ can be expressed as%
\begin{align}
\left\langle \mathcal{H}\right\rangle _{G}  &  =\sum_{i\alpha,j\beta,\sigma
}\left(  z_{i\alpha\sigma}z_{j\beta\sigma}t_{i\alpha j\beta}+\varepsilon
_{i\alpha}\delta_{\alpha\beta}\delta_{ij}\right)  \left\langle c_{i\alpha
\sigma}^{\dagger}c_{j\beta\sigma}\right\rangle _{0}\nonumber\\
&  +\sum_{i\mathcal{F}}U_{i\mathcal{F}}p_{i\mathcal{F}}%
\end{align}
where $\left\langle c_{i\alpha\sigma}^{\dagger}c_{j\beta\sigma}\right\rangle
_{0}=\left\langle \Psi_{0}\left\vert c_{i\alpha\sigma}^{\dagger}%
c_{j\beta\sigma}\right\vert \Psi_{0}\right\rangle $ and $U_{i\mathcal{F}%
}=\left\langle \mathcal{F}_{i}\left\vert \mathcal{H}_{1}\right\vert
\mathcal{F}_{i}\right\rangle $. We define the Gutzwiller orbital
renormalization factor $z_{i\alpha\sigma}\equiv1$ for Ba 5d and As 4p orbitals
in the ddp-model since they are not considered as strongly correlated
orbitals. Following Ref. \cite{Bunemann07}, for the Fe/Mn 3d orbitals%
\begin{equation}
z_{i\gamma\sigma}=\frac{1}{\sqrt{n_{i\gamma\sigma}^{0}\left(  1-n_{i\gamma
\sigma}^{0}\right)  }}\sum_{\mathcal{F},\mathcal{F}^{\prime}}\sqrt
{p_{i\mathcal{F}}p_{i\mathcal{F}^{\prime}}}\left\vert \left\langle
\mathcal{F}_{i}\left\vert c_{i\gamma\sigma}^{\dagger}\right\vert
\mathcal{F}_{i}^{\prime}\right\rangle \right\vert ^{2}%
\end{equation}
with $n_{i\gamma\sigma}^{0}=\left\langle n_{i\gamma\sigma}\right\rangle _{0}$.
$p_{i\mathcal{F}}$ is the occupation probability of configuration $\left\vert
\mathcal{F}_{i}\right\rangle $, which absorbs $g_{i\mathcal{F}}$ and serves as
the additional variational parameter due to the Gutzwiller wave function.

Since $\mathcal{H}_{0}$\ in the Hamiltonian of Eq. \ref{Hamil} is obtained by
downfolding the DFT band structure, some contributions from $\mathcal{H}_{1}$
have already been taken into account by $\mathcal{H}_{0}$\ in a mean-field
way. Such contributions are commonly referred to as the double counting term.
Consequently, the total energy of the system is given by%
\begin{equation}
E_{tot}=\left\langle \mathcal{H}\right\rangle _{G}+E_{d.c.}%
\end{equation}
with%
\begin{equation}
E_{d.c.}=-\sum_{i}\left(  \bar{U}N_{id}\left(  N_{id}-1\right)  /2-\bar
{J}N_{id}\left(  N_{id}/2-1\right)  /2\right)
\end{equation}
following the treatment in DFT+U method\cite{DC1,DC2}. Here $\bar{U}$
($\bar{J}$) is the averaged Coulomb U (Hund's rule coupling J) parameter,
which for d orbitals is given as\cite{Dai09,Kotliar06}%
\begin{equation}
\bar{U}=\frac{U+4U^{\prime}}{5}%
\end{equation}%
\begin{equation}
\bar{J}=\bar{U}-U^{\prime}+J
\end{equation}
$N_{id}=\sum_{\gamma\sigma}n_{i\gamma\sigma}^{0}$ is the expectation value of
the total Fe/Mn d-orbital occupancy at site $i$. The double counting term
$E_{d.c.}$ introduces some spin-independent chemical potential shift for the
Fe/Mn 3d orbitals,%
\begin{equation}
\mathcal{\tilde{\varepsilon}}_{i\gamma}=-\bar{U}\left(  N_{id}-\frac{1}%
{2}\right)  +\bar{J}\left(  N_{id}/2-\frac{1}{2}\right)
\end{equation}
In the Fe/Mn d-model that contains only 3d states, $\mathcal{\tilde
{\varepsilon}}_{i\gamma}$\ has no effect since it merely shifts all the Fe/Mn
3d levels by a constant. However, in the ddp-models $\mathcal{\tilde
{\varepsilon}}_{i\gamma}$\ becomes important because it changes the position
of the Fe/Mn 3d levels with respect to Ba 4d and As 4p orbitals and hence
affects the electron flow between these orbitals.

The minimization of $E_{tot}$\ with the constraints of $\sum_{\mathcal{F}%
}p_{i\mathcal{F}}=1$ and $\sum_{\mathcal{F}}p_{i\mathcal{F}}n_{i\gamma\sigma
}^{\mathcal{F}}=n_{i\gamma\sigma}^{0}$\ yields an effective single-particle
Hamiltonian%
\begin{equation}
H_{\text{eff}}=\sum_{i\alpha,j\beta,\sigma}\left(  z_{i\alpha\sigma}%
z_{j\beta\sigma}t_{i\alpha j\beta}+\mu_{i\alpha\sigma}\delta_{\alpha\beta
}\delta_{ij}\right)  c_{i\alpha\sigma}^{\dagger}c_{j\beta\sigma} \label{Heff}%
\end{equation}
where $\mu_{i\alpha\sigma}=\varepsilon_{i\alpha}+\mathcal{\tilde{\varepsilon}%
}_{i\alpha}+\frac{\partial z_{i\alpha\sigma}}{\partial n_{i\alpha\sigma}^{0}%
}e_{i\alpha\sigma}+\eta_{i\alpha\sigma}$ if $\alpha$\ belongs to the
correlated Fe/Mn 3d orbitals and $\mu_{i\alpha\sigma}=\varepsilon_{i\alpha}%
$\ otherwise.\ It has been shown that $H_{\text{eff}}$ describes the
Landau-Gutzwiller quasiparticle bands\cite{Bunemann03}, and the square of the
orbital renormalization factor $z^{2}$ corresponds to the quasiparticle
weight\cite{Dai09}. $\left\{  z_{i\gamma\sigma}\right\}  $ and $\left\{
\eta_{i\gamma\sigma}\right\}  $\ are obtained by solving the following
equations.%
\begin{equation}
\sum_{\mathcal{F}^{\prime}}\mathcal{M}_{\mathcal{FF}^{\prime}}^{i}%
\sqrt{p_{i\mathcal{F}^{\prime}}}=\eta_{i0}\sqrt{p_{i\mathcal{F}}} \label{Meff}%
\end{equation}
where
\begin{equation}
\mathcal{M}_{\mathcal{F},\mathcal{F}^{\prime}}^{i}=\sum_{\gamma\sigma}%
\frac{e_{i\gamma\sigma}}{2\sqrt{n_{i\gamma\sigma}^{0}\left(  1-n_{i\gamma
\sigma}^{0}\right)  }}\left\vert \left\langle \mathcal{F}_{i}\left\vert
c_{i\gamma\sigma}^{\dagger}+c_{i\gamma\sigma}\right\vert \mathcal{F}%
_{i}^{\prime}\right\rangle \right\vert ^{2}+\delta_{\mathcal{FF}^{\prime}%
}\left(  U_{i\mathcal{F}}-\sum_{\gamma\sigma}\eta_{i\gamma\sigma}%
n_{i\gamma\sigma}^{\mathcal{F}}\right)
\end{equation}
with $e_{i\gamma\sigma}=\sum_{j\beta}\left(  z_{j\beta\sigma}t_{i\alpha
j\beta}\left\langle c_{i\gamma\sigma}^{\dagger}c_{j\beta\sigma}\right\rangle
_{0}+c.c.\right)  $. $\Psi_{0}$ from Eq. \ref{Heff} and $\left\{
p_{i\mathcal{F}}\right\}  $ from Eq. \ref{Meff} need to be solved
self-consistently to obtain the final solution\cite{YYX11}.

\subsection{LDA+Gutzwiller}

The \bigskip above formalism based on the Gutzwiller approximation on the
multiband model can be naturally combined with DFT, in which full electron
density convergence can be achieved\cite{Dai08,Dai09,Ho08}. The total energy
density functional can be written as%
\begin{equation}
E\left[  \rho\right]  =\left\langle \Psi_{G}\left\vert \hat{T}\right\vert
\Psi_{G}\right\rangle +E_{H}\left[  \rho\right]  +E_{xc}+\int\rho\left(
\mathbf{r}\right)  v_{ion}\left(  \mathbf{r}\right)  d^{3}r+E_{ion-ion}%
\end{equation}
where $\hat{T}$\ is the kinetic energy operator and $E_{H}$\ is the Hartree
potential energy. The electron-ion and ion-ion potential energies are given by
$\int\rho\left(  \mathbf{r}\right)  v_{ion}\left(  \mathbf{r}\right)  d^{3}r$
and $E_{ion-ion}$, respectively. In the LDA+Gutzwiller
method\cite{Dai08,Dai09}, the approximate exchange-correlation energy $E_{xc}$
takes the following form%
\begin{equation}
E_{xc}=E_{xc}^{LDA}\left[  \rho\right]  +\sum_{i\mathcal{F}}U_{i\mathcal{F}%
}p_{i\mathcal{F}}+E_{d.c.}%
\end{equation}

The electron density $\rho$\ under the Gutzwiller approximation can be written
as%
\begin{equation}
\rho\left(  \mathbf{r}\right)  =\sum_{i\alpha,j\beta,\sigma}^{\prime
}z_{i\alpha\sigma}z_{j\beta\sigma}\varphi_{i\alpha}^{\ast}\left(
\mathbf{r}\right)  \varphi_{j\beta}\left(  \mathbf{r}\right)  \left\langle
c_{i\alpha\sigma}^{\dagger}c_{j\beta\sigma}\right\rangle _{0}+\sum
_{i\alpha,\sigma}\left\vert \varphi_{i\alpha}\left(  \mathbf{r}\right)
\right\vert ^{2}\left\langle n_{i\alpha\sigma}\right\rangle _{0}%
\end{equation}
where $\left\{  \varphi_{i\alpha}\right\}  $ is a general complete basis set
including local correlated orbitals (e.g., Fe/Mn 3d orbitals) as well as
nonlocal orbitals. $\sum^{\prime}$ indicates that the summation excludes the
term with $i\alpha=j\beta$. The kinetic energy is also renormalized%
\begin{equation}
\left\langle \Psi_{G}\left\vert \hat{T}\right\vert \Psi_{G}\right\rangle
=\sum_{i\alpha,j\beta,\sigma}^{\prime}z_{i\alpha\sigma}z_{j\beta\sigma
}\left\langle \varphi_{i\alpha}\left\vert \hat{T}\right\vert \varphi_{j\beta
}\right\rangle \left\langle c_{i\alpha\sigma}^{\dagger}c_{j\beta\sigma
}\right\rangle _{0}+\sum_{i\alpha,\sigma}\left\langle \varphi_{i\alpha
}\left\vert \hat{T}\right\vert \varphi_{i\alpha}\right\rangle \left\langle
n_{i\alpha\sigma}\right\rangle _{0}%
\end{equation}

The total energy $E\left[  \rho\right]  $ of the closed form can be similarly
minimized as in the above multiband model with respect to $\Psi_{0}$\ and
$\left\{  p_{i\mathcal{F}}\right\}  $. The resultant effective single-particle
Hamiltonian has the same form as Eq. \ref{Heff} with the hopping element
\begin{equation}
t_{i\alpha j\beta}=\left\langle \varphi_{i\alpha}\left\vert \hat{T}\right\vert
\varphi_{j\beta}\right\rangle +\int\left(  v_{scr}^{LDA}\left(  \mathbf{r}%
\right)  +v_{ion}\left(  \mathbf{r}\right)  \right)  \varphi_{i\alpha}^{\ast
}\left(  \mathbf{r}\right)  \varphi_{j\beta}\left(  \mathbf{r}\right)  d^{3}r
\end{equation}
The LDA screening potential $v_{scr}^{LDA}\left(  \mathbf{r}\right)  $ is the
sum of the Hartree potential $v_{H}\left(  \mathbf{r}\right)  =\frac{\delta
E_{H}\left[  \rho\right]  }{\delta\rho}$ and the exchange-correlation
potential $v_{xc}^{LDA}\left(  \mathbf{r}\right)  =\frac{\delta E_{xc}%
^{LDA}\left[  \rho\right]  }{\delta\rho}$. The renormalized orbital level
$\mu_{i\alpha\sigma}=\varepsilon_{i\alpha}+\tilde{\varepsilon}_{i\alpha}%
+\frac{\partial z_{i\alpha\sigma}}{\partial n_{i\alpha\sigma}^{0}}\tilde
{e}_{i\alpha\sigma}+\eta_{i\alpha\sigma}$ if $\alpha$\ belongs to the
correlated Fe/Mn 3d orbitals and $\mu_{i\alpha\sigma}=\varepsilon_{i\alpha}%
$\ otherwise. Here $\varepsilon_{i\alpha}=t_{i\alpha i\alpha}$ and
$e_{i\gamma\sigma}=\sum_{j\beta}\left(  z_{j\beta\sigma}t_{i\alpha j\beta
}\left\langle c_{i\gamma\sigma}^{\dagger}c_{j\beta\sigma}\right\rangle
_{0}+c.c.\right)  .\left\{  z_{i\gamma\sigma}\right\}  $ and $\left\{
\eta_{i\gamma\sigma}\right\}  $\ are obtained by solving the same equation as
Eq.\ref{Meff}. One can see that the major difference is that the hopping
element $t_{i\alpha j\beta}$\ and the orbital level $\varepsilon_{i\alpha}%
$\ will be updated self-consistently by the LDA screening potential
$v_{scr}^{LDA}\left(  \mathbf{r}\right)  $\ in LDA+Gutzwiller method, while
they are fixed in the multiband model.

\section{Effects caused by downfolding}%

\begin{figure}
[ptb]
\begin{center}
\includegraphics[
height=2.2175in,
width=3.2011in
]%
{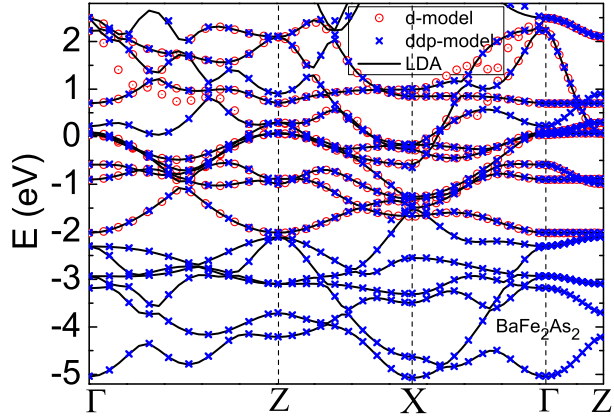}%
\caption{(Color online) Band structure of BaFe$_{2}$As$_{2}$ in PM\ phase
based on LDA (solid line), Fe d-model (circle) and ddp-model calculations
(cross). Fermi level is shifted to zero.}%
\label{bfabd}%
\end{center}
\end{figure}
\begin{figure}
[ptb]
\begin{center}
\includegraphics[
height=2.2024in,
width=3.2011in
]%
{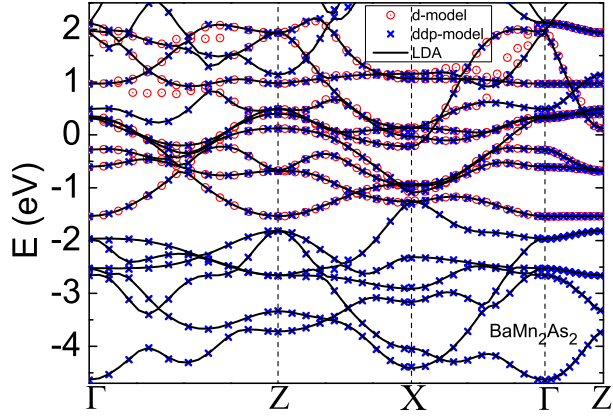}%
\caption{(Color online) Band structure of BaMn$_{2}$As$_{2}$ with settings
same as Fig. \ref{bfabd}.}%
\label{bmabd}%
\end{center}
\end{figure}
\begin{figure}
[ptb]
\begin{center}
\includegraphics[
height=6.2496cm,
width=6.2496cm
]%
{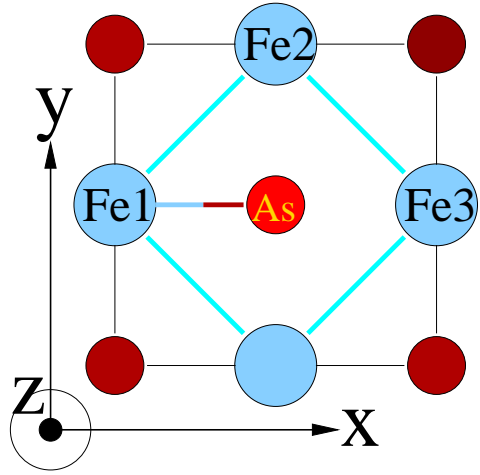}%
\caption{(Color online) The atomic configuration of the Fe-As layer with the
coordinate system. Note that the center As atom is above the Fe-plane, while
the corner As atoms are below the Fe-plane.}%
\label{stru}%
\end{center}
\end{figure}
\begin{table}[tbp] \centering
\resizebox{\textwidth}{!}{\begin{tabular}
[c]{|c|c|c|c|c|c|}\hline
\textbf{Fe}$_{1}$/Mn$_{1}$ & $xy$ & $yz$ & $z^{2}$ & $xz$ & $x^{2}-y^{2}$\\\hline
Fe$_{2}$/Mn$_{2}$: $xy$ & \textbf{-0.34}/-0.32 (\textbf{-0.35}/-0.33) &
\textbf{0.25}/0.20 (\textbf{0.21}/0.17) & \textbf{0.31}/0.28 (\textbf{0.27}/0.24) & \textbf{-0.25}/-0.20 (\textbf{-0.21}/-0.17) & 0\\\hline
$yz$ & \textbf{0.25}/0.20 (\textbf{0.21}/0.17) & \textbf{-0.22}/-0.16
(\textbf{-0.14}/-0.10) & \textbf{-0.11}/-0.10 (\textbf{-0.09}/-0.08) &
\textbf{0.12}/0.11 (\textbf{0.08}/0.08) & \textbf{0.18}/0.16 (\textbf{0.17}/0.15)\\\hline
$z^{2}$ & \textbf{0.31}/0.28 (\textbf{0.27}/0.24) & \textbf{-0.11}/-0.10
(\textbf{-0.09}/-0.08) & \textbf{0.06}/0.03 (\textbf{0.01}/-0.01) &
\textbf{0.11}/0.10 (\textbf{0.09}/0.08) & 0\\\hline
$xz$ & \textbf{-0.25}/-0.20 (\textbf{-0.21}/-0.17) & \textbf{0.12}/0.11
(\textbf{0.08}/ 0.08) & \textbf{0.11}/0.10 (\textbf{0.09}/0.08) &
\textbf{-0.22}/-0.16 (\textbf{-0.14}/-0.10) & \textbf{0.18}/0.16
(\textbf{0.17}/0.14)\\\hline
$x^{2}-y^{2}$ & 0 & \textbf{0.18}/0.16 (\textbf{0.17}/0.15) & 0 &
\textbf{0.18}/0.16 (\textbf{0.17}/0.15) & \textbf{-0.13}/-0.07 (\textbf{-0.09}/-0.03)\\\hline
Fe$_{3}$/Mn$_{3}$: $xy$ & \textbf{-0.07}/-0.05 (\textbf{-0.06}/-0.05) &
\textbf{0.14}/0.11 (\textbf{0.10}/0.08) & 0 & 0 & 0\\\hline
$yz$ & \textbf{-0.14}/-0.11 (\textbf{-0.10}/-0.08) & \textbf{0.15}/0.11
(\textbf{0.08}/0.07) & 0 & 0 & 0\\\hline
$z^{2}$ & 0 & 0 & \textbf{-0.01}/-0.02 (\textbf{0}/-0.01) & \textbf{0.17}/0.15
(\textbf{0.13}/0.12) & \textbf{-0.18}/-0.13 (\textbf{-0.12}/-0.08)\\\hline
$xz$ & 0 & 0 & \textbf{-0.16}/-0.15 (\textbf{-0.13}/-0.12) & \textbf{0.32}/0.27 (\textbf{0.20}/0.17) & \textbf{0.04}/0.06 (\textbf{0.04}/0.05)\\\hline
$x^{2}-y^{2}$ & 0 & 0 & \textbf{-0.17}/-0.13 (\textbf{-0.12}/-0.08) &
\textbf{-0.04}/-0.06 (\textbf{-0.04}/-0.05) & \textbf{0.11}/0.09
(\textbf{0.09}/0.07)\\\hline
As: $y$ & (\textbf{-0.50}/-0.42) & (\textbf{0.39}/0.33) & 0 & 0 & 0\\\hline
$z$ & 0 & 0 & (\textbf{0.34}/0.29) & (\textbf{0.09}/0.12) & (\textbf{-0.49}/-0.41)\\\hline
$x$ & 0 & 0 & (\textbf{0.14}/0.13) & (\textbf{-0.69}/-0.58) & (\textbf{0.32}/0.25)\\\hline
\end{tabular}
}\caption{The hopping parameters between the orbitals of Fe1/Mn1 and those of its nearest neighbour Fe2/Mn2, next nearest neighbour Fe3/Mn3 and the center As atom. The entries of \textbf{A1}/B1(\textbf{A2}/B2) mean that \textbf{A1} is the hopping parameter in the Fe d-model and B1 in the Mn d-model. \textbf{A2} and B2 are the corresponding values in the ddp-models.}\label{tbp}%
\end{table}%

The Fe/Mn d-model and ddp-model for BaFe$_{2}$As$_{2}$\ and BaMn$_{2}$As$_{2}%
$\ have been obtained by downfolding the first-principles LDA electron bands
of the experimentally determined crystal structures\cite{Rotter081,Singh091}
using the quasi-atomic minimal basis-set orbitals
(QUAMBOs)\cite{QUAMBO1,QUAMBO2,QUAMBO3,QUAMBO4}. The main idea of the QUAMBO
approach is to recover a set of local quasi-atomic orbitals by performing an
inverse unitary transformation for the \textquotedblleft
bonding\textquotedblright\ and \textquotedblleft
anti-bonding\textquotedblright\ states. In practice, the \textquotedblleft
bonding states\textquotedblright\ are some occupied bands or low energy bands
which are intended to be preserved in the new local orbital representation.
The \textquotedblleft anti-bonding\textquotedblright\ states are constructed
from the orthogonal subspace of the \textquotedblleft
bonding\textquotedblright\ subspace and are optimized such that resultant
QUAMBOs have maximal similarity with the free atomic orbitals. The orthogonal
subspace is spanned by the wave functions in some energy window. It becomes
complete in the infinite energy or band limit, where succinct closed form for
QUAMBOs can still be obtained\cite{QUAMBO3}. In principle, different choices
of the \textquotedblleft bonding\textquotedblright\ \ states and the energy
window for the \textquotedblleft anti-bonding\textquotedblright\ states will
yield different set of QUAMBOs and tight-binding parameters, which is
reasonable and acceptable since the downfolding of the band structure should
not be unique. We find that the ddp model will give too many Fe/Mn d-electrons
if it is obtained by treating all the occupied states and virtual states up to
$2$eV above Fermi level as the \textquotedblleft bonding\textquotedblright%
\ states in the infinite band limit: $7.3$ d-electrons per Fe atom for
BaFe$_{2}$As$_{2}$ (close to the analysis of Ref.\cite{Ciraci09}) and $6.2$
for BaMn$_{2}$As$_{2}$, while the nominal value would be $6$ and $5$,
respectively. Such large deviation of the correlated electron number clearly
do not properly reflect the correct physics of these materials. This problem
does not exist for the Fe/Mn d-model where the electron filling is fixed by
assumption. In order to have a general tight-binding model suitable for the
correlated local orbital-based approaches, we choose a finite energy window
around the Fermi level for the generation of the \textquotedblleft
anti-bonding\textquotedblright\ states and another smaller energy window to
control the Bloch bands which will contribute to the Fe/Mn d orbitals. We can
then construct tight-binding models with reasonable number of correlated
d-electrons. As a result, we have $6.3$ Fe 3d-electrons and $5.3$ Mn
3d-electrons for the ddp-model. Those are the initial occupancies that are
further reduced due to the self-consistent determination of the occupancies
within the Gutzwiller approach and the double counting corrections, reaching
values close to $6$ Fe 3d-electrons and $5$ Mn 3d-electrons.

Figure \ref{bfabd} and \ref{bmabd}\ show the bare band structures ($U=J=0$) of
the d-model and the ddp-model for BaFe$_{2}$As$_{2}$\ and BaMn$_{2}$As$_{2}$,
respectively. The LDA band structures have also been plotted for comparison.
One can see that the band structure from the ddp-model agrees very well with
the LDA\ result in the low energy window for both systems. The Fe/Mn d-model,
however, fails to reproduce one low energy unoccupied band near $\Gamma$-point
which is mainly contributed by the Ba 5d orbitals. The total number of
electrons is $6(5)$ per Fe(Mn) atom for the d-model and $12(11)$ per Fe(Mn)
atom for the ddp-model. The tight-binding parameters between one Fe/Mn atom
and its nearest Fe/Mn atom, second the nearest Fe/Mn atom and the nearest As
atom are listed in Table \ref{tbp}, with the atomic configuration and the
coordinate system shown in Fig.\ref{stru}. The hopping parameters of the
BaFe$_{2}$As$_{2}$ system are overall larger in magnitude than those of
BaMn$_{2}$As$_{2}$ system, which is mainly owing to the fact that BaMn$_{2}%
$As$_{2}$\ has a much larger volume than BaFe$_{2}$As$_{2},$ although the
ionic radii of Fe and Mn are very close. The direct Fe-Fe/Mn-Mn hopping
parameters become smaller in the ddp-model, as expected. The Fe d-model has
been compared with the one downfolded using maximally localized Wannier
function (MLWF)\cite{Marzari97,Souza01}. Although the MLWF approach can
achieve better fitting after some fine tuning\cite{Graser10}, we find that the
results reported here are not sensitive to this detail.%
\begin{figure}
[ptb]
\begin{center}
\includegraphics[
height=1.8245in,
width=3.186in
]%
{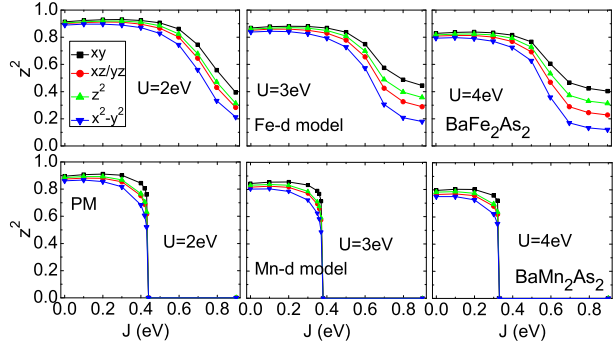}%
\caption{(Color online) The square of the orbital renormalization factor
$z_{i\gamma}^{2}$\ as a function of Hund's rule coupling J parameter at
Hubbard U=2eV, 3eV\ and 4eV in the Fe d-model (upper panels) for BaFe$_{2}%
$As$_{2}$\ and Mn d-model for BaMn$_{2}$As$_{2}$ (lower pannels) in the
paramagnetic state.}%
\label{pmz2d}%
\end{center}
\end{figure}
\begin{figure}
[ptb]
\begin{center}
\includegraphics[
height=2.0489in,
width=3.5923in
]%
{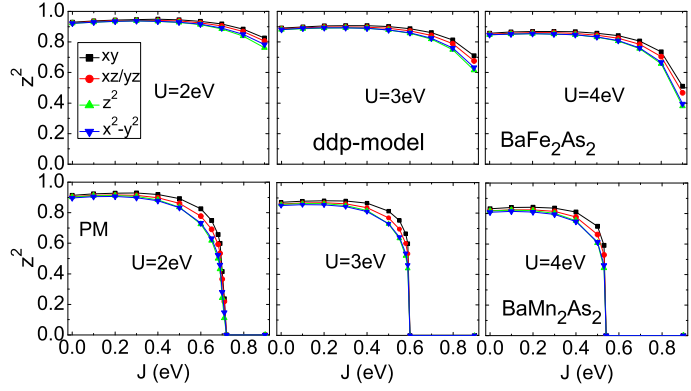}%
\caption{(Color online) Same as Fig. \ref{pmz2d} but based on ddp-model.}%
\label{pmz2ddp}%
\end{center}
\end{figure}

\section{Comparative study of the d and ddp model for PM phase}

Figure \ref{pmz2d} and \ref{pmz2ddp}\ show the variation of the $z^{2}%
$-factors for the correlated d-orbitals with increasing Hund's rule coupling
J\ and Hubbard U\ fixed at 2eV, 3eV\ and 4eV in the paramagnetic state of the
Fe/Mn d-model and the ddp-model, respectively. On the small J side, the
$z^{2}$-factors for both models of the compounds stay almost the same. With
increasing J, the $z^{2}$-factors start to drop rapidly but remain finite for
the two models of BaFe$_{2}$As$_{2}$. In contrast, the $z^{2}$-factors exhibit
a sharp decrease to zero for both models of BaMn$_{2}$As$_{2}$, i.e., Mott
localization for all the orbitals sets in beyond a threshold value of $J$.
This indicates that BaMn$_{2}$As$_{2}$ is a more correlated system than
BaFe$_{2}$As$_{2}$. By increasing $U$ from $2$eV\ to $4$eV, the transition of
the $z^{2}$-curves occurs at somewhat smaller $J$-values\ for all the model
calculations. The results show a much stronger dependence on the Hund's rule
coupling $J$\ rather than Hubbard $U$, which is typically the case in
correlated multiband systems. Comparing the Fe/Mn d-model and the ddp-model,
we observe a systematic shift of the $z^{2}$-curves to larger $J$\ value with
the inclusion of the Ba 5d and As 4p orbitals to the Fe/Mn d-model. Thus, a
larger value of $J$ is needed in a ddp-model to obtain the similar $z^{2}%
$-factors as in the Fe/Mn d-model. This can be attributed to the screening
effect of additional electrons in the ddp-model as caused the hybridization of
the Fe/Mn 3d orbitals with other, less correlated degrees of freedom.
\begin{figure}
[ptb]
\begin{center}
\includegraphics[
height=1.9934in,
width=3.5924in
]%
{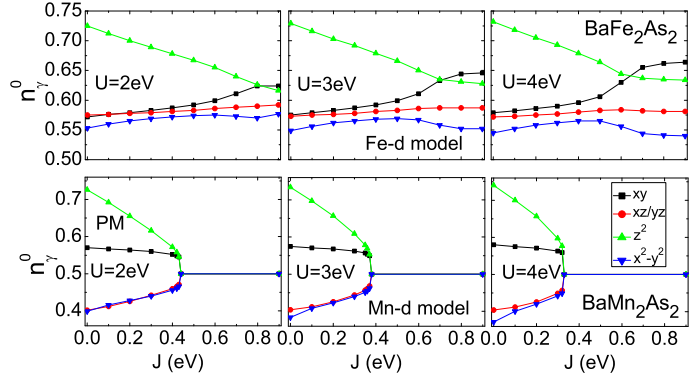}%
\caption{(Color online) Orbital occupations for the Mn/Fe d-model with
settings same as Fig. \ref{pmz2d}.}%
\label{pmnd}%
\end{center}
\end{figure}
\begin{figure}
[ptb]
\begin{center}
\includegraphics[
height=1.9655in,
width=3.5923in
]%
{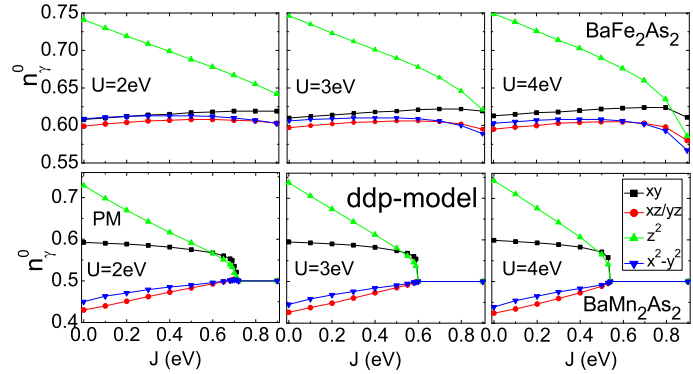}%
\caption{(Color online) Local Fe/Mn 3d rbital occupations for the ddp-model
with settings same as Fig. \ref{pmz2d}.}%
\label{pmnddp}%
\end{center}
\end{figure}

\begin{figure}
[ptb]
\begin{center}
\includegraphics[
height=5.7046cm,
width=8.1012cm
]%
{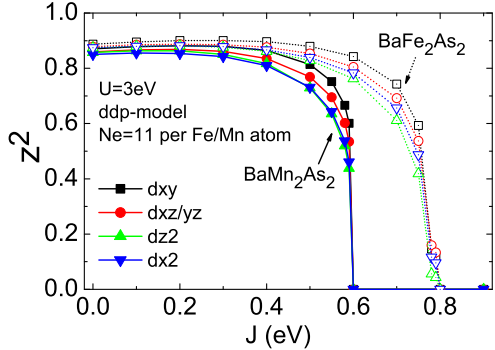}%
\caption{(Color online) The square of the orbital renormalization factor
$z_{i\gamma}^{2}$\ as a function of Hund's rule coupling $J$ parameter for the
ddp-model of BaMn$_{2}$As$_{2}$ (solid symbol) and that of BaFe2As2 (open
symbol) with total number of electrons fixed to 11 per Fe/Mn atom and U=3eV.}%
\label{pmz2ddpbmafechk}%
\end{center}
\end{figure}

The general feature of the local d-orbital occupations in the model
calculations as shown in Fig. \ref{pmnd} and \ref{pmnddp} is consistent with
that of the $z^{2}$-factors described above. While the Hubbard U\ parameter
tends to polarize the orbitals, the Hund's rule coupling J favors the
equalization of the local orbital occupations. All local d-orbitals become
half occupied as the system approaches the Mott transition for the models of
BaMn$_{2}$Fe$_{2}$ which has 5 electrons in each 5 Mn 3d orbitals. For the
BaFe$_{2}$As$_{2}$ system, there is 6 electrons in each 5 Fe 3d orbitals.
Hence we expect an orbital selective Mott transition would occur with large
$U$ and $J$\cite{Anisimov02,Koga04}. We did perform model calculations at even
larger $U$ and find an orbitally selective transition where the $z$ factors of
all orbitals except the $d_{xy}$-orbital vanish, while $z_{xy}^{2}=1$ as the
orbital becomes completely filled. This behavior requires however unphysically
large values of $U$. Indeed some signature can be identified as one can see
that $n_{x^{2}-y^{2}}^{0}$ becomes closest to 0.5 and $z_{x^{2}-y^{2}}^{2}$
drops most rapidly with increasing $U$\ and $J$ for the Fe d-model. When
comparing the results of Mn/Fe d-model and the ddp-model, a similar systematic
shift of the local orbital occupation curves to the large $J$\ side is
observed due to the screening and hybridization effect. We conclude that
BaMn$_{2}$Fe$_{2}$ is approaching a Mott insulator with large localized
moment. Combining our calculations in the magnetically ordered state and in
the paramagnetic state we obtain $J\simeq0.2-0.3$eV to yields the
experimentally observed ordered moment, values that are not yet in the Mott
insulating regime, yet they are rather close. In contrast, BaFe$_{2}$As$_{2}$
is significantly away from Mott localization, yet there are clearly visible
polarizations of the orbital populations that demonstrate that the system is
in an intermediate regime. To demonstrate that this is indeed the case we
determined the quasiparticle weight of a system with the tight binding
parameters (ddp-model) of BaFe$_{2}$As$_{2}$, yet with a total electron count
of 11 per Fe atom as appropriate for BaMn$_{2}$As$_{2}$ (see Fig.
\ref{pmz2ddpbmafechk}). Now a Mott transition similar to the actual BaMn$_{2}%
$As$_{2}\ $calculation occurs, albeit at a somewhat larger value of $J$.%

\section{Effects of charge self-consistency}%

\begin{figure}
[ptb]
\begin{center}
\includegraphics[
height=1.8245in,
width=3.186in
]%
{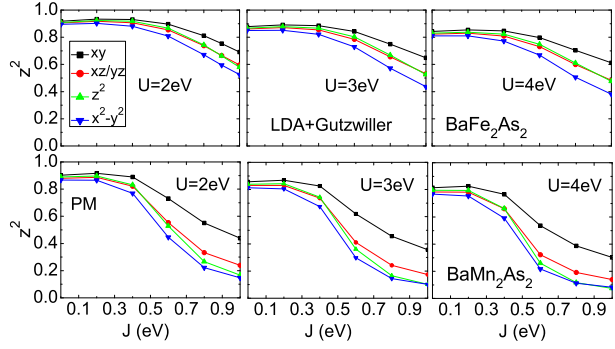}%
\caption{(Color online) Same as Fig. \ref{pmz2d} but based on LDA+G
calculations.}%
\label{pmz2ldag}%
\end{center}
\end{figure}
\begin{figure}
[ptb]
\begin{center}
\includegraphics[
height=1.7403in,
width=3.2002in
]%
{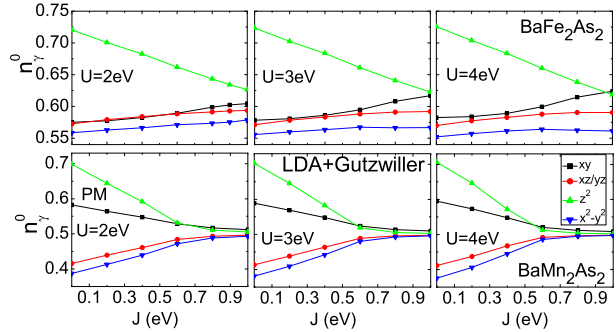}%
\caption{(Color online) Local Fe/Mn 3d orbital occupations based on
LDA+G\ calculations with settings same as Fig. \ref{pmz2d}.}%
\label{pmnldag}%
\end{center}
\end{figure}

The variation of the $z^{2}$-factors and occupations of the correlated Fe/Mn
3d orbitals with U\ and J\ has further been investigated by the
LDA+Gutzwiller\ method\cite{Dai08,Dai09}, which is built upon the
LDA+U\ approach\cite{Anisimov91,Anisimov97}. We use the same correlated
orbitals obtained via the QUAMBO procedure as in the Fe/Mn d-models for the
calculations and achieve the electron density self-consistency. As shown in
Fig. \ref{pmz2ldag}, the $z^{2}$-curves exhibit a similar trend but vary in a
way much slower than that in the Fe/Mn d-models, which can again be attributed
to the screening effect of the other electrons in the systems. Our results on
the BaFe$_{2}$As$_{2}$ system are in good agreement with those reported in
Ref. \cite{Dai10}. Overall, the $z^{2}$-factors for BaMn$_{2}$As$_{2}$ are
systematically smaller than those for BaFe$_{2}$As$_{2}$. Note that the
calculated screened U(J)\ is 3.6(0.76)eV for Fe and 3.2(0.70)eV for Mn based
on the constrained random phase approximation and the maximally localized
Wannier function\cite{Miyake08}, it is fairly reasonable to assume that the
screened interaction parameters in the models investigated here are also very
close. Therefore BaMn$_{2}$As$_{2}$ is a more correlated material than
BaFe$_{2}$As$_{2}$.

\section{Comparative study of the d and ddp model for AFM phase}%

\begin{figure}
[ptb]
\begin{center}
\includegraphics[
height=2.6663in,
width=3.1993in
]%
{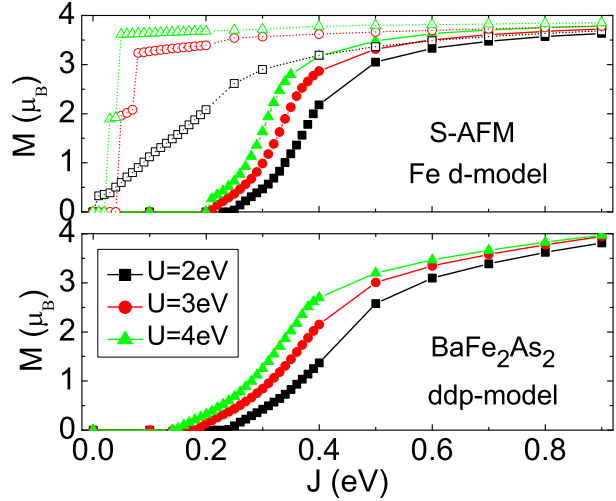}%
\caption{(Color online) Magnetic moment vs J\ at U=2eV (squares), 3eV
(circles) and 4eV (triangles) in the Fe d-model (upper panel) and the
ddp-model (lower panel) for BaFe$_{2}$As$_{2}$. The solid symbols indicate the
results using Gutzwiller approximation, and open symbols using Hartree-Fock
approximation.}%
\label{bfaafmm}%
\end{center}
\end{figure}

\begin{figure}
[ptb]
\begin{center}
\includegraphics[
height=2.6787in,
width=3.2002in
]%
{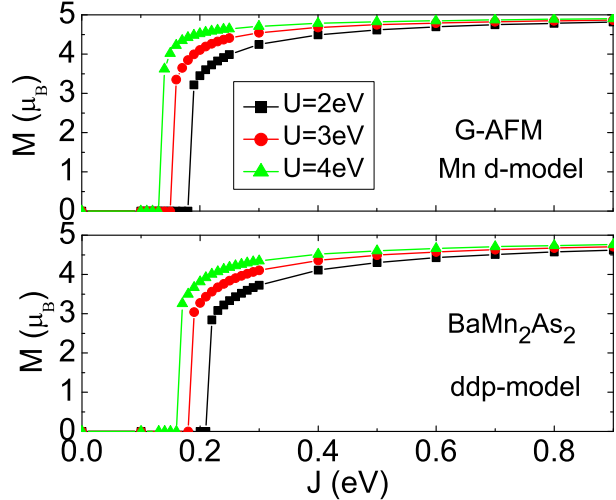}%
\caption{(Color online) Magnetic moment vs J\ curves for BaMn$_{2}$As$_{2}$
with setting as as Fig. \ref{bfaafmm}.}%
\label{bmaafmm}%
\end{center}
\end{figure}

The ground state AFM\ phase has also been studied for the models of the two
systems. Figure \ref{bfaafmm} shows the behavior of the magnetic moment as a
function of J\ with U\ fixed at 2eV, 3eV\ and 4eV in the stripe-type AFM
ground state for BaFe$_{2}$As$_{2}$. In contrast to the Fe d-model including
complete local interactions for LaOFeAs in which the system exhibits a sharp
transition from PM state to AFM state with large magnetic moment using the
Gutzwiller approximation\cite{Bune11}, the models with only density-density
type interactions for BaFe$_{2}$As$_{2}$ describe the PM-AFM\ transition
fairly well as the magnetic moment increases rapidly yet smoothly when
J\ rises as indicated by the solid symbols in the upper panel. The magnetic
moment varies from $0.14\mu_{B}$ to $1.38\mu_{B}$ in the range of
$J=0.22-0.32$eV with $U=3$eV. For reference the experimental magnetic moment
is $0.87\mu_{B}$ for BaFe$_{2}$As$_{2}$\cite{Huang08}. With the inclusion of
the Ba 5d and As 4p orbitals in the ddp-model, the magnetic moment increases
slower as seen in the lower panel. The high sensitivity of the magnetic moment
to Hund's rule coupling J\ is consistent with the DFT+DMFT calculation
results\cite{Yin11}. For comparison, Fig. \ref{bfaafmm} also shows the
magnetic moment of the AFM\ phase in the Fe d-model with Hartree-Fock
approximation by the open symbols in the upper panel. Within Hartree-Fock
approximation, the PM-AFM transition takes place at much smaller J. The
magnetic moment exhibits quite smooth variation with increasing J\ at U=2eV.
However, an abrupt transition from PM\ to AFM\ state with large magnetic
moment occurs at U=3eV. In comparison, the models of BaMn$_{2}$As$_{2}$ yield
a very sharp transition from PM\ to the G-type AFM\ ground state with magnetic
moment $M\approx3\mu_{B}$ as shown in Fig. \ref{bmaafmm}. However, it is still
reasonable since the experimental magnetic moment is 3.88$\mu_{B}%
$\cite{Singh091}. The $z^{2}$-factors for the models of BaFe$_{2}$As$_{2}$ in
the AFM\ ground state are always or order unity ($>0.8$).%

\section{Conclusion}

In summary, we have studied the Fe/Mn d-model and the ddp-model on the
electronic and magnetic properties of the BaFe$_{2}$As$_{2}$ and BaMn$_{2}%
$As$_{2}$ systems. The renormalization factors of the correlated Mn 3d
orbitals are found to be systematically smaller than those of Fe 3d orbitals,
implying that the electron correlation for BaMn$_{2}$As$_{2}$ are much more
efficient to cause Mott localization physics compared to BaFe$_{2}$As$_{2}$.
Ultimately this is due to the fact that Fe, with its close to six electrons,
must undergo an orbital selective Mott transition while the odd number of
electrons in case of Mn allow for a more ordinary Mott transition. While the
strength of the interactions are not sufficient for Mott localization in
either system the Mn-based material is significantly closer to localization.
The LDA+Gutzwiller results on the paramagnetic phase with electron density
self-consistency also confirm the conclusion. The variation of the magnetic
moment in the AFM ground state of the two compounds seems in accordance with
the experimental results: a smooth increase of the magnetic moment with rising
$J$ for the stripe-AFM\ ground state of BaFe$_{2}$As$_{2}$ with small
experimental magnetic moment and an abrupt jump of the magnetic moment from
$0$ to about $3\mu_{B}$ for the G-AFM\ ground state of BaFe$_{2}$As$_{2}$ with
large experimental magnetic moment. \ We also checked that the two different
ordered states are indeed energetically lower in the respective systems. The
Gutzwiller approximation, however, is not able to provide a quantitatively
consistent description of the AFM\ ground state for the iron pnictide systems
with both correct magnetic moment and band renormalization factor for the same
parameters. Nevertheless it is a comparatively easy to implement and powerful
tool that allows for a first analysis of the role of magnetic correlations,
the nature of magnetically ordered states in\ strongly correlated
multi-orbital materials.

\begin{acknowledgments}
This work was supported by the U.S. Department of Energy, Office of Basic
Energy Science, Division of Materials Science and Engineering including a
grant of computer time at the National Energy Research Supercomputing Center
(NERSC) at the Lawrence Berkeley National Laboratory under Contract No. DE-AC02-07CH11358.
\end{acknowledgments}

\end{document}